\documentclass{ws-procs9x6}

\newcommand{\beq}{\begin{equation}}
\newcommand{\eeq}{\end{equation}}
\newcommand{\bea}{\begin{eqnarray}}
\newcommand{\eea}{\end{eqnarray}}
\newcommand{\bmp}{\begin{minipage}}
\newcommand{\emp}{\end{minipage}}
\newcommand{\D}{\displaystyle}

\newcommand{\tr}{{\rm tr}}

\begin{document}

\title{Connecting short to long scales in the Confining Vacuum}

\author{E.~T. TOMBOULIS\footnote{\uppercase{W}ork partially
supported by grant NSF-PHY-0309362.}}

\address{Department of Physics and Astronomy \\
University of California, Los Angeles\\
CA 90095-1547, USA \\ 
E-mail: tombouli@physics.ucla.edu}


\maketitle

\abstracts{
We study approximate decimations in SU(N) LGT that connect the 
short to long distance regimes. Simple  
`bond-moving' decimations turn out to provide both upper 
and lower bounds on the exact partition function. 
This leads to a representation of the exact partition function in  terms of 
successive decimations whose effective couplings flows are related to 
those of the easily computable bond-moving decimations. The implications 
for a derivation of confinement  
from first principles are discussed.}

\section{Introduction}

Over the last several years an enormous amount of work has been 
performed by lattice workers on the physics of the QCD vacuum. 
In particular, isolating the types of configurations in 
the functional measure that are responsible for maintaining 
confinement at (arbitrarily) weak coupling has been a central  
issue. A great deal of 
information concerning the confinement mechanism   
has been obtained from these investigations (for recent review, 
see\cite{Gr}). 
However, the goal of a direct derivation of confinement from first principles 
has remained elusive for the last thirty years. 

The origin of the difficulty is clear. One is faced with 
a multi-scale problem involving the passage from the short-distance   
weakly coupled, ordered regime to the long distance strongly 
coupled, disordered, confining regime. 
Such variable multi-scale behavior can only be addressed by some 
nonperturbative block-spinning or decimation procedure capable of 
bridging these different regimes. Exact 
decimation schemes appear analytically hopeless, and numerically very 
difficult. It is not even clear what a good definition of block-spin 
variables would be. There is, however, a class of approximate simple  
decimation procedures which are known in many cases to give qualitatively  
correct results. They are generally known as `bond moving' decimations. 
Here we will consider such decimations in a somewhat more general form 
and show that they can provide bounds on the exact theory.  
This leads to a representation of the partition function of 
the exact theory which allows a connection to be made to the behavior of 
the approximate, but easily computable, decimations at 
successive length scales. The implications for 
the question of an actual derivation of confinement in LGT will be 
discussed below.  

The framework applies to general $SU(N)$, though explicit numerical 
or analytical calculations supporting the considerations below have 
for the most part been carried out only for $SU(2)$. 
    
\section{Bond moving decimations} 

Starting with some plaquette action, e.g the Wilson action 
$A_p(U) ={\beta\over N}\;{\rm Re}\,\tr U_p$, at lattice spacing $a$, 
we consider the character expansion of the exponential of the action: 
\beq 
F(U, a) =e^{A_p(U)} 
   = \sum_j\;F_j(\beta,a)\,d_j\,\chi_j(U) \label{exp}
\eeq
with Fourier coefficients: 
\beq F_j = \int\,dU\;F(U,a)\,{1\over d_j}\,\chi_j(U)\; .
\eeq
Here $\chi_j$ denotes the character of the $j$-th representation 
of dimension $d_j$. $j=0$ will always denote the trivial representation. 
E.g, for SU(2), $j=0, {1\over 2}, 1, {3\over 2}, \ldots$, and $d_j=(2j+1)$.
In terms of normalized coefficients: 
\beq 
c_j = {\D F_j \over \D F_0} \;,
\eeq
one then has  
\bea
F(U, a) &=&  F_0\,\Big[\, 1 + \sum_{j\not= 0} c_j(\beta)\,d_j\,\chi_j(U)\,
 \Big] \nonumber \\
    & \equiv & F_0\;f(U,a)  \label{nexp} 
\eea 
For a reflection positive action one necessarily has: 
\beq
F_j \geq 0\qquad \mbox{hence}\quad 1\geq c_j\geq 0 \qquad\quad 
\mbox{all}\quad j \;.
\eeq
The partition function on lattice $\Lambda$ is then  
\beq
Z_\Lambda(\beta) 
                    =F_0^{|\Lambda|}\; \int dU_\Lambda\;\prod_p\,f_p(U,a)
\;.\label{PF1}
\eeq

\begin{figure}[ht]
\centerline{\epsfxsize=3in\epsfbox{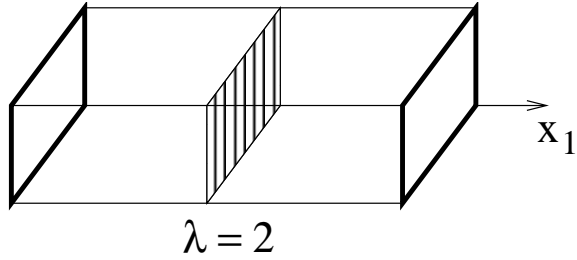}}   
\caption{Basic bond (plaquette) moving operation. \label{dec1}}
\end{figure}

\begin{figure}[ht]
\centerline{\epsfxsize=3in\epsfbox{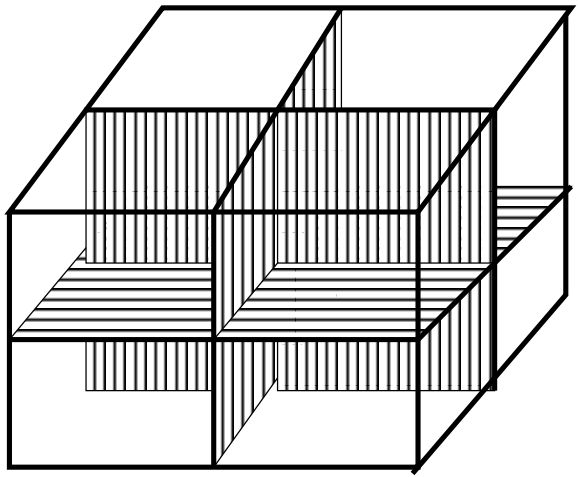}}   
\caption{Isotropic interior plaquette moving operation in a 
decimation cell (hypercube). \label{dec2}}
\end{figure}

We now consider RG decimation 
transformations $a \to \lambda a$ in, say, the $x^1$-direction 
(Figure~\ref{dec1}).   
Simple approximate transformations of the `bond moving' type 
are implemented by  
`weakening', i.e. decreasing the $c_j$'s  of interior plaquettes (shaded), 
and `strengthening', i.e. increasing $c_j$'s  of boundary plaquettes 
(bold) in every decimation cell of side length $\lambda$. 
The simplest scheme\cite{MK}, which is  
adopted in the following,  implements  
complete removal, $c_j=0$, of interior plaquettes. This  
is performed simultaneously in all directions (Figure~\ref{dec2}). 

Under successive decimations 
\bea 
& & a \to \lambda a \to \lambda^2 a \to \cdots \to \lambda^n a \nonumber \\
 & & \Lambda \to \Lambda^{(1)} \to \Lambda^{(2)} \to \cdots \to 
\Lambda^{(n)} \nonumber  
\eea
the RG transformation rule is then:
\beq
f(U,n-1)\to f(U,n) = \Big[ 1 + \sum_{j\not= 0} c_j(n)\,d_j
\,\chi_j(U) \Big] \label{RG1}
\eeq
with:  
\beq
c_j(n) = F_j(n)/F_0(n)\, , \quad F_j(n)=
\Big[\hat{F}_j(n)\Big]^{\lambda^2} , \label{RG2}
\eeq   
\beq
\hat{F}_j(n)=  \int\,dU\;\Big[\,f(U,n-1)\,\Big]^\nu\,{1\over d_j}\,\chi_j(U) 
\; . \label{RG3}
\eeq
The parameter $\nu$ controls by how much the remaining plaquettes 
have been strengthened to compensate for the removed plaquettes.  
What has been considered in the literature before is 
$\nu=\lambda^{(d-2)}$, where $d$ is the spacetime dimension. This 
choice of $\nu$ defines the MK recursions\cite{MK}. Here we 
generalize to consider $\nu$ an arbitrary parameter. 
 
The resulting partition function after $n$ decimation steps is:  
\beq 
Z_\Lambda(\beta, n) 
                    =\prod_{m=0}^n F_0(m)^{|\Lambda|/\lambda^{md}}\; 
\int dU_{\Lambda^{(n)}}\;\prod_p\,f_p(U,n) \;.\label{PF2}
\eeq

It is important to note that after each decimation step 
the resulting action retains the 
original {\it one-plaquette} form but will, in general,  
contain all representations:
\beq
A_p(U,n)= \sum_j \; \tilde{\beta}_j(\beta)\,\chi_j(U) \;.
\eeq 
Furthermore, among the effective couplings 
$\tilde{\beta}_j$ some negative ones may in general occur. 
These features are present even after a single decimation 
step $a\to \lambda a$ starting with the usual single 
representation (fundamental) Wilson action.  

Preservation of the one-plaquette form of the action is of course what 
makes these decimations simple to explore. The rule specified by (\ref{RG1})-
(\ref {RG3}) is meaningful for any real (positive) $\nu$. 
Here, however, a basic distinction can be made. 
For {\it integer} $\nu$, 
the important property of positivity of the Fourier 
coefficients in (\ref{exp}), (\ref{nexp}):   
\beq
F_0(n)\geq 0 \quad , \qquad c_j(n)\geq 0 \;, 
\eeq 
and hence reflection positivity are maintained at each decimation step. 
This, in general, is {\it not} the case for non-integer $\nu$. 
Thus non-integer $\nu$ results in approximate RG transformations that 
violate the reflection positivity of the theory (assuming a 
reflection positive starting action).\footnote{It is worth 
noting in this context that numerical investigations 
of the standard MK recursions, at least for gauge theories, appear to 
have been carried out for the most part for fractional $\lambda$, 
($1< \lambda <2$), which corresponds to non-integer $\nu$; e.g. 
see\cite{NT,BGZ}.} 

There are various other interesting features of such decimations. 
The following property, in particular, is  
important. Define a normalized $\hat{F}_j(n)$ (cf. (\ref{RG3})):
\beq
\hat{c}_j(n)\equiv \hat{F}_j(n)/\hat{F}_0(n) \leq 1 , 
\qquad \mbox{so that} 
\qquad c_j(n) = \hat{c}_j(n)^{\lambda^2} \;. \label{nF}
\eeq
Then it is possible to prove that 
\beq 
\sum_j \;c_j(n)\,\big(\,\hat{c}_j(n+1) - c_j(n)\,\big) \geq 0 \;.
\label{cineq1}
\eeq
It follows from (\ref{cineq1}) that the norm ($l_2$ norm) of 
the vector formed from the $\hat{c}_j(n+1)$ coefficients is bigger  
than that of the vector of the $c_j(n)$. 
In fact one finds in explicit numerical evaluations that 
(\ref{cineq1}) holds component-wise, i.e. $\hat{c}_j(n+1) \geq c_j(n)$. 

As can be seen from the relation between $\hat{c}_j(n+1)$ and $c_j(n+1)$ 
in (\ref{nF}), (\ref{RG3}), however, it can still be 
that the norm of the $c_j(n+1)$'s is smaller than that of the $c_j(n)$'s. 
i.e. the norm of the normalized coefficients $c_j(n)$ in (\ref{RG1}) 
decreases under successive decimations. Note, in particular, that 
when $\nu$ is taken to depend on $\lambda$, $d$, the resulting highly 
nonlinear dependence can give very nontrivial behavior. 
This is in fact what happens in the case of the MK  
recursions where $\nu=\lambda^{(d-2)}$: the normalized coefficients 
$c_j(n)$ do decrease under successive decimations
in the approach to a fixed point in lower dimensions. But an upper 
critical dimension arises above which  
the $\hat{c}_j(n+1)$'s become sufficiently large compared to the $c_j(n)$'s 
so that this is no longer the case, and triviality ensues (for the RG 
flow on the weak coupling side).

\section{The exact partition function}
Since our decimations are not exact decimation transformations, the partition 
function does not in general remain invariant under them. 
The subsequent development hinges on the following two basic 
statements that can now be proved: 

(I) With $\nu=\lambda^{d-2}$:  
\beq
Z_\Lambda(\beta, n) \leq Z_\Lambda(\beta, n+1) \;.\label{I}
\eeq 

(II) With $\nu=1$: 
\beq
Z_\Lambda(\beta, n+1) \leq Z_\Lambda(\beta, n) \; .\label{II}
\eeq 
Note that for  $d=2$ (\ref{I}) - (\ref{II}) express the well-known 
fact that the decimations become exact. For $d>2$, in both (I), (II) one 
in fact has strict inequality. 

(I) says that modifying the couplings of the remaining plaquettes after 
decimation by taking $\nu=\lambda^{d-2}$ (standard MK choice\cite{MK}) 
results into overcompensation (upper bound on the partition function). 
Translation invariance and convexity of the free energy as a function of the 
couplings in the action underlie (\ref{I}).  

(II) says that decimating plaquettes while leaving the couplings of the 
remaining plaquettes unaffected results in a lower bound on the partition  
function. Reflection positivity (positivity of Fourier coefficients) 
is crucial for this to hold.

Consider now the, say, $(n-1)$-th decimation step  
with Fourier coefficients $c_j(n-1)$, which we relabel   
$c_j(n-1)=\tilde{c}_j(n-1)$.  
Given these $\tilde{c}_j(n-1)$, we proceed to compute the coefficients 
$F_0(n)$, $c_j(n)$ of the next decimation step 
according to (\ref{RG1})-(\ref{RG3}) above with $\nu=\lambda^{d-2}$. 
 
Then introducing a parameter $\alpha$, ($0\leq \alpha$), define the 
{\it interpolating coefficients}:
\beq
\tilde{c}_j(n,\alpha)= \tilde{c}_j(n-1)^{\lambda^{2}(1-\alpha)}\,
 c_j(n)^\alpha \, . \label{inter1}
\eeq
Then, 
\beq
\tilde{c}_j(n,\alpha)= \left\{ \begin{array}{lll}
c_j(n) & : &\alpha=1 \\ 
\tilde{c}_j(n-1)^{\lambda^{2}} & : & \alpha=0 
\end{array} \right. \label{inter2}
\eeq
The $\alpha=0$ value is that of the $n$-th step coefficients 
resulting from (\ref{RG1})-(\ref{RG3}) with $\nu=1$.

Thus defining  the corresponding partition function
\bea    
Z_\Lambda(\beta, \alpha, n) &= & 
\big(\prod_{m=0}^{n-1} F_0(m)^{|\Lambda|/\lambda^{md}}\big)\;
F_0(n)^\alpha \nonumber \\ 
& & \cdot\;
\int dU_{\Lambda^{(n)}}\;\prod_p\,f_p(U,n,\alpha) \label{PF3}
\eea
where 
\beq
f_p(U,n,\alpha)= \Big[ 1 + \sum_{j\not= 0} \tilde{c}_j(n,\alpha)
\, d_j\,\chi_j(U) \Big] \, ,
\eeq
we have from (\ref{I}), (\ref{II})), and (\ref{inter2}) above: 
\beq
Z_\Lambda(\beta, 0, n) \leq Z_\Lambda(\beta, n-1) 
\leq Z_\Lambda(\beta, 1, n)  \;. \label{III} 
\eeq
Now the partition function (\ref{PF3}) is a continuous, 
in fact analytic, in $\alpha$. So (\ref{III}) implies that, by continuity, 
there exist a value of $\alpha$: 
\[ \alpha=\alpha^{(n)}(\beta, \lambda, \Lambda)\,, \qquad  
0 < \alpha^{(n)}(\beta,\lambda, \Lambda) < 1  \]  
such that  
\beq 
Z_\Lambda(\beta, \alpha^{(n)}, n)= Z_\Lambda(\beta, n-1) \;.
\eeq
In other words there is an $\alpha$ at which the 
$n$-th decimation step partition function equals that obtained at 
the previous decimation step; the partition function does not 
change its value under the decimation step $\lambda^{n-1} a 
\to \lambda^n a$.

So starting at original spacing $a$, at every decimation step 
$m$, ($m=0,1,\cdots,n$), there exist a value $0< \alpha^{(m)} <1$ 
such that 
\beq
Z_\Lambda(\beta, \alpha^{(m+1)}, m+1)=
Z_\Lambda(\beta, \alpha^{(m)}, m)\, .
\eeq 

This then gives, after $n$ successive decimations, an {\it exact 
representation} of the original partition function in the form:
\bea    Z_\Lambda(\beta) &= & 
F_0^{|\Lambda|}\; \int dU_\Lambda\;\prod_p\,f_p(U,a) \nonumber \\
  & =& \prod_{m=0}^n F_0(m)^{\alpha^{(m)}|\Lambda|/\lambda^{md}}\;
\nonumber \\ 
& & \cdot\;
\int dU_{\Lambda^{(n)}}\;\prod_p\,f_p(U,n,\alpha^{(n)}) , \label{rep} 
\eea
i.e. in terms of the successive bulk free energy contributions from the 
$a \to \lambda \to \cdots \to \lambda^n a$ decimations 
and a one-plaquette effective action on the resulting 
lattice $\Lambda^{(n)}$. 

The $\alpha^{(m)}$'s in (\ref{rep})  
may be viewed as effective couplings which,   
in addition to the $\{\tilde{\beta}_j^{(m)}\}$, enter in 
the specification of the effective action and bulk free energy 
at each decimation step $m$. Thus the flow from scale $a$ 
to scale $\lambda^n a$ is now specified by  
$\{ \mbox{\boldmath$\tilde{\beta}$}(n, \beta, \lambda ),
\; \mbox{\boldmath$\alpha$}(n, \beta, \lambda, \Lambda)\}
\equiv  \{\; \{\tilde{\beta}_j^{(m)}(\beta, \lambda )\}\,, 
\alpha^{(m)}(\beta, \lambda, \Lambda) \;|\;m=0,\ldots ,n 
\;\}$. This dependence on the additional couplings  
\boldmath$\alpha$ \unboldmath  may be considered as compensating 
for the absence in (\ref{rep}) of additional terms, beyond 
the one-plaquette interaction, that would normally be 
expected in an effective action. 

At weak and strong coupling $\alpha^{(m)}$ may be estimated analytically. 
At large $\beta$, where the decimations approximate the free energy 
rather accurately, the appropriate $\alpha$ values are very close to 
unity. At strong coupling they may be estimated by comparison with 
the strong coupling expansion. On any finite lattice there is also a  
weak volume dependence as a correction which goes away    
as an inverse power of the lattice size. 

For most purposes the exact values of the $\alpha^{(m)}$'s, 
beyond the fact that are fixed between $0$ and $1$, are not 
immediately relevant. The main point of the representation (\ref{rep}) is 
that it can in principle relate the behavior of the exact theory to 
that of (modifications of) the easily computable approximate 
decimations.

Indeed, starting from the $\tilde{c}_j(n-1,\alpha^{(n-1)})$'s at the   
$(n-1)$-th  step, consider the coefficients at the next step, and compare   
those evaluated at $\alpha=\alpha^{(n)}$, i.e.   
\ $\tilde{c}_j(n,\alpha=\alpha^{(n)})$, to  
those evaluated at $\alpha=1$, i.e   
\ $\tilde{c}_j(n,\alpha=1)\equiv c_j(n)$. The latter  
will be referred to as 
the MK coefficients. (Recall that $\alpha=1 \Longleftrightarrow \nu=
\lambda^{d-2}$, the standard MK choice. The absence 
of a tilde on a coefficient in the following 
always means that it is computed 
at $\alpha=1$.) According to (I), the MK coefficients give 
an upper bound. 

To facilitate the comparison let us rewrite (\ref{inter1}) in the 
form 
\beq 
\tilde{c}_j(n,\alpha)= \left(\,{\tilde{c}_j(n-1)
\over \hat{c}_j(n)}\,\right)^{\lambda^{2}(1-\alpha)}\,
 c_j(n)\, . \label{inter3}
\eeq 
Now property (\ref{cineq1}) and the remark following it imply that the ratio 
in the brackets in (\ref{inter3}) is less or equal to unity. 
It follows that 
\beq
\tilde{c}_j(n,\alpha) \leq c_j(n) \qquad \mbox{for any} 
\quad 0\leq \alpha\leq 1 \;.\label{compineq}
\eeq 
This has the following important consequence. 

Assume we are in a dimension $d$ such that 
under successive decimations the MK coefficients ($\alpha=1$) 
are non-increasing. Then (\ref{compineq}) implies: 
\bea
  \tilde{c}_j(n,\alpha^{(n)}) & \geq &  c_j(n+1) \geq  
\tilde{c}_j(n+1,\alpha^{(n+1)}) \nonumber \\
     & \geq & c_j(n+2) \geq 
\tilde{c}_j(n+2,\alpha^{(n+2)}) \nonumber \\
     &\geq & \cdots    \nonumber 
\eea

Thus, if  the $c_j(n)$'s are non-increasing, 
so are the $\tilde{c}_j(n,\alpha)$. 
The  $c_j(n)$'s must then approach a fixed point, and hence so must the 
$\tilde{c}_j(n,\alpha)$'s, since $c_j(n), \tilde{c}_j(n,\alpha)\geq 0$. 
Note the fact that this conclusion is independent of the specific 
value of the $\alpha$'s at every decimation step.

In particular, if the $c_j(n)$'s approach the strong coupling  
fixed point, i.e. $F_0\to 1$, $c_j(n) \to 0$ as $n\to \infty$, so must 
the $\tilde{c}_j(n,\alpha)$'s of the exact representation. 
If the MK decimations confine, so do those in the exact 
representation (\ref{rep}). 
As it is well-known by explicit numerical evaluation, the MK decimations 
for $SU(2)$ and $SU(3)$ indeed confine  
for all $\beta<\infty$ and $d\leq 4$. Above the critical dimension 
$d=4$, the decimations result in free spin wave behavior.  

\section{Discussion and outlook}
What do the above results say about the question of confinement in 
the exact theory? They are clearly strongly suggestive of confinement 
for all $\beta$ in the exact theory. They cannot, however, as yet be 
taken to constitute an actual proof.
The statement at the end of the previous section concerns the behavior of 
the long distance action part in 
the representation (\ref{rep}). Now  (\ref{rep}) also includes the 
large free energy bulk contributions 
from integration over all scales from $a$ to $\lambda^n a$.     
It is the complete representation involving both contributions 
that provides an equality to the 
value of the exact partition function.  This, just 
by itself, does not suffice to rigorously isolate, 
at least in any direct way, 
the actual behavior of the corresponding 
long distance part in the exact theory. To do this one needs to consider 
order parameters which can directly couple to the corresponding long 
distance parts of the effective action in the exact theory and any 
representation of it like that given by (\ref{rep}). In other words, one 
would need to  carry through the above derivation given 
for the partition function 
also for the case of appropriate order parameters.   

The derivation of the basic two statements (I) and (II) above 
(eqs. (\ref{I}), (\ref{II})) assumes translation invariance and 
reflection positivity. In the presence of observables 
such as a Wilson loop, translation invariance is broken and 
reflection positivity is reduced to hold only in the plane bisecting the loop. 
This does not allow the above derivations to be carried through in any 
obvious way. Fortunately, there are other order parameters that can  
characterize the possible phases of the theory while  
maintaining  translational invariance. They are the well-known vortex 
free energy, and its $Z(N)$ Fourier transform (electric flux free energy). 
They are in fact the natural order parameters in the present context 
since they are constructed out of partition functions.
Recall that the vortex free energy is defined by 
\beq 
e^{-F_{v}(\tau)} = Z_\Lambda(\tau)/Z_\Lambda \; .\label{vfe} 
\eeq 
Here $Z_\Lambda(\tau)$ denotes the partition function  with action 
modified by the `twist' $\tau \in Z(N)$ for every plaquette on a 
coclosed set of plaquettes $V$ winding through the 
periodic lattice in $(d-2)$ directions; e.g. for the Wilson action 
one has the replacement ${\beta\over N}\;{\rm Re}\,\tr U_p  \to 
{\beta\over N}\;{\rm Re}\,\tr U_p\tau$ for every $p \in V$.   
The twist represents a discontinuous 
gauge transformation on the set $V$ which introduces 
$\pi_1(SU(N)/Z(N))$ vortex flux rendered topologically 
stable by being wrapped around the lattice torus.  
As indicated by the notation, 
$Z_\Lambda(\tau)$ depends only on the presence of the flux, and is 
invariant under changes in the exact location of $V$. 
The vortex free energy is then 
the ratio of the partition function in the presence of this external 
flux to the partition function in the 
absence of the flux (the latter is what was considered above). 
The above development, in particular the derivation of (\ref{rep}), 
should then be repeated also for $Z_\Lambda(\tau)$. Bulk (local) 
free energy contributions resulting from integrating over successive 
scales are insensitive to the presence of the flux. Thus in the 
analog to (\ref{rep}) for $Z_\Lambda(\tau)$ only the long distance 
effective action part would be affected by its presence, and the 
bulk contributions would cancel in (\ref{vfe}). 
Statements, as the ones obtained in the previous section, concerning the 
behavior of the long distance parts in such 
representations of the two factors in (\ref{vfe}) would 
then directly constrain the corresponding behavior in the exact theory.   

There is, however, an immediate technical complication in obtaining the analog 
to (\ref{rep}) for $Z_\Lambda(\tau)$. The presence of the flux  
reduces reflection positivity to hold only in planes 
perpendicular to the  directions in which 
the flux winds through the lattice. The simple nature of the decimations, 
however, makes it plausible that this still suffices to 
allow a generalization of the previous derivation for $Z_\Lambda$ to 
go through also in the case of $Z_\Lambda(\tau)$.  
Further investigation of this and related questions 
will be reported elsewhere.  

\section*{Acknowledgments} 
I would like to thank the participants of Confinement 2003 for many 
discussions. I am grateful to Prof. H. Suganuma and the organizing 
committee for the invitation 
and for organizing such a stimulating and enjoyable conference.


\begin{thebibliography}{9}
\bibitem{Gr} J. Greensite, Progr. Part. Nucl. Phys. {\bf 51}, 1 (2003), 
             (hep-lat/0301023).  
\bibitem{MK} A. A. Migdal, Sov. Phys. JETP {\bf 42}, 413, 743  (1976); 
             L. Kadanoff, Ann. Phys. (N.Y.) {\bf 100}, 369 (1976).
\bibitem{NT} M. Nauenberg and D. Toussaint, Nucl. Phys. {\bf B190} [FS3], 
             217 (1981). 
\bibitem{BGZ} K. Bitar, S. Gottlieb and C. Zachos, Phys. Rev. {\bf D26}, 
             2853 (1982). 
\end{thebibliography}
\end{document}